# Thermodynamique Quantique et Perspectives Quantiques

# Quantum Thermodynamics and Quantum Perspectives

Camille LOMBARD LATUNE

*Université Bourgogne Europe, CNRS, Laboratoire Interdisciplinaire Carnot de Bourgogne ICB UMR 6303, F-21000 Dijon, France*

**RÉSUMÉ.** Après une brève perspective historique, nous introduisons les notions incontournables de travail et de chaleur pour les systèmes quantiques, pour ensuite aborder les moteurs quantiques fonctionnant sur les cycles d'Otto et de Carnot quantiques. Le caractère irréversible et dissipatif du cycle d'Otto quantique est brièvement analysé ainsi que les raisons physiques sous-jacentes, contrastant avec l'optimalité énergétique du cycle de Carnot quantique. La question centrale des effets quantiques est aussi abordée et accompagnée de plusieurs exemples. Finalement, la dernière partie s'efforce d'expliquer le rôle que la thermodynamique quantique joue pour les applications quantiques et les technologies quantiques, notamment par rapport à l'optimisation énergétique et au compromis entre performances et coûts énergétiques.

**ABSTRACT.** After a brief historical perspective, we introduce the key notions of work and heat for quantum systems, to then apply them to quantum engines operating on quantum Otto and Carnot cycles. The irreversible and dissipative character of the quantum Otto cycle is briefly analyzed, contrasting with the energetic optimality of the quantum Carnot cycle. The central question of quantum effects is also addressed and illustrated with several examples. Finally, the last part strives to explain the role that quantum thermodynamics plays for quantum applications and quantum technologies, particularly in relation to energy optimization and the trade-off between performances and energy costs.

**MOTS-CLÉS.** Thermodynamique ; systèmes quantiques ; qubits ; cycle de Carnot ; irréversibilité ; avantages quantiques ; fluctuations quantiques ; compromis entre performances et coûts énergétiques.

## 1. Introduction

L'objectif et la motivation de cette communication sont de présenter un bref panorama du domaine en plein essor qu'est la Thermodynamique Quantique. Nous commencerons par une rapide perspective historique du domaine suivie par un bref aperçu de l'évolution et de l'état actuel de la Thermodynamique Quantique.

### 1.1. Perspective historique

Le premier article abordant explicitement certaines thématiques chères à ce qui deviendra des années plus tard la Thermodynamique Quantique, est l'article de 1959 de Scovil et Schulz-DuBois [1] qui propose une version du maser (microwave amplified stimulated emission radiation, l'équivalent du laser dans le domaine des micro-ondes) à la manière d'un moteur thermique. Quelques années plus tard, avec l'aide de Geusic, ils formalisent leur idée initiale et proposent un modèle quantique [2] du fameux cycle de Carnot [3]. Une décennie plus tard, dans une approche un peu plus fondamentale, Spohn propose une définition pour la production d'entropie associée à un système quantique ouvert [4] et pour l'irréversibilité qui en découle [5]. Dans la foulée, Alicki publie un article appliquant la théorie des Systèmes Quantiques Ouverts et moteurs thermiques [6], suivi par Kosloff [7], qui publie aussi un autre papier [8] sur les aspects thermodynamiques de la mesure en Mécanique Quantique.

### 1.2. La diversification de la Thermodynamique Quantique

Malgré une apparente intensification des publications au début des années 80, il faudra attendre le début des années 2000 pour voir l'essor de la Thermodynamique Quantique. Néanmoins, ces quelques articles constituent la base et l'origine de la thermodynamique quantique, et ont introduit des problématiques qui restent chères au domaine jusqu'à aujourd'hui. Cependant, le domaine s'est beaucoup diversifié au fil des années, couvrant de nombreuses thématiques comme : avantages quantiques pour les machines thermiques quantiques et pour les batteries quantiques ; extraction de travail et "ergotropy" ; machines quantiques produisant des ressources quantiques (cohérences et intrication quantiques) ; élaboration d'un formalisme thermodynamique dans un contexte quantique généralisé ; production d'entropie et origine quantique de l'irréversibilité ; thermodynamique quantique en couplage fort ; thermodynamique quantique autonome ; thermodynamique quantique stochastique ; théorème des fluctuations ; contrôle des fluctuations ; raccourcis pour l'évolution adiabatique ("shortcut-to-adiabaticity") ; thermométrie quantique ; thermalisation en système quantique fermé ; théorie des ressources ; vitesse limite quantique ("quantum seep limit") ; informatique quantique thermodynamique ; énergétique quantique ; horloges quantiques. Ces nombreuses thématiques visent à répondre à plusieurs grandes questions et motivations de la Thermodynamique Quantique actuelle, comme entre autres, le concept de travail et de chaleur à l'échelle quantique dans un contexte général ; l'origine de l'irréversibilité ainsi que la production d'entropie dans un contexte quantique général, le compromis « performance des opérations quantiques » *versus* « coût énergétique », le coût énergétique du traitement de l'information, et finalement la compréhension, puis le contrôle des fluctuations quantiques.

Comme il est impossible de tout couvrir, nous allons nous concentrer seulement sur certains aspects spécifiques. Néanmoins, il est important de garder à l'esprit que la Thermodynamique Quantique regroupe en fait un large ensemble de thématiques, comme mentionné plus haut. S'inspirant de l'émergence historique de la Thermodynamique Quantique, nous allons commencer par nous pencher sur les moteurs thermiques quantiques, ce qui implique en amont de comprendre et définir les concepts de travail et chaleur. C'est aussi un point d'entrée naturel pour les non-spécialistes. Nous mentionnons en particulier certains effets quantiques et les perspectives expérimentales. Puis nous continuons en abordant l'importante question des fluctuations quantiques, et nous terminons avec une rapide présentation du rôle croissant de la Thermodynamique Quantique dans le développement des technologies quantiques.

## 2. Moteurs thermiques quantiques

### *2.1. Le travail quantique*

La notion de travail pour des systèmes quantiques est inspirée de la notion classique, qui découle de la définition de travail en Mécanique Classique, $W = \vec{F}.\vec{\Delta r}$, où $\vec{F}$ représente la force appliquée au système considéré (système mécanique, ou ensemble de particules, ou même particule unique), et $\vec{\Delta r}$ représente le déplacement de ce dernier. En thermodynamique stochastique classique, le travail peut aussi s'exprimer en termes de l'Hamiltonien $H(\lambda;q,p)$ (classique) du système, sous la forme suivante [9] :

$$W = \int_0^t du \dot{\lambda}(u) \frac{\partial H(\lambda; q, p)}{\partial \lambda},$$

où (q,p) dénote les coordonnées généralisées du système, et λ(t) représente une trajectoire dans l'espace des paramètres (typiquement l'intensité d'un champ de contrôle) de l'Hamiltonien. Cette dernière forme motive la définition quantique suivante, pour un échange de travail infinitésimal,

$$\delta W = \dot{W} dt := \text{Tr}[\rho(t)\dot{\hat{H}}(t)]dt,$$

et pour un travail fini, on intègre dans le temps le travail infinitésimal,

$$W = \int_0^t du \text{Tr}[\rho(u)\dot{\hat{H}}(u)],$$

où ρ(u) représente l'état quantique du système quantique considéré, Ĥ(u) son Hamiltonien (en Mécanique Quantique, l'Hamiltonien est un opérateur qui agit sur l'espace vectoriel qui décrit les états du système quantique), et le symbole **Tr** signifie la « trace », qui veut dire que l'on calcule la moyenne de l'Hamiltonien Ĥ(u) par rapport à l'état quantique ρ(u). À noter que, avec cette définition, on retrouve que le travail correspond à de l'énergie injectée par un système extérieur ou mécanique, comme pour la thermodynamique classique. Cela vient du fait qu'en Mécanique Quantique la dépendance temporelle de l'Hamiltonien d'un système quantique provient d'un contrôle extérieur agissant sur ce même système.

*2.2. La chaleur quantique*

Par opposition au travail, la chaleur pour un système quantique correspond au transfert d'énergie lié au changement d'état du système quantique. En terme mathématique, pour un échange infinitésimal de chaleur, cela s'exprime comme,

$$\delta Q = \dot{Q} dt := \text{Tr}[\dot{\rho}(t)\hat{H}(t)]dt,$$

et pour un échange fini, on intègre dans le temps l'expression précédente

$$Q = \int_0^t du \text{Tr}[\dot{\rho}(u)\hat{H}(u)].$$

Avec cette définition, on peut montrer que la partie unitaire de la dynamique ne contribue pas au transfert de chaleur. On retrouve donc, comme pour le travail, l'essence de la définition classique : l'échange de chaleur provient de la partie non-unitaire de la dynamique, c'est-à-dire par exemple de l'interaction du système avec un *bain thermique* (expression qui désigne l'ensemble des degrés de libertés externes interagissant avec le système). Observons que les deux définitions précédentes de W et Q sont valides dans un contexte de couplage faible entre le ou les systèmes quantiques et leurs bains, ainsi que l'hypothèse que ces bains sont dans des états thermiques et sont beaucoup plus grands que les systèmes quantiques auxquels ils sont couplés. Par contre, en l'absence de ces hypothèses, dans un contexte plus général, il existe de nombreuses définitions alternatives suivant la situation analysée, mais ce sont des questions de recherche actuelle sur lesquelles il n'y a pas encore de consensus.

*2.3. Le cycle d'Otto quantique*

Le cycle d'Otto quantique suit quatre étapes adaptées du cycle d'Otto classique [10]. Pour permettre une description simplifiée, nous allons considérer un cycle pour un qubit réalisé par un spin-1/2, c'est-à-dire un système quantique possédant deux niveaux d'énergie, un état fondamental et un état excité. Le cycle utilise deux types d'action sur le spin. La première est de modifier le gap d'énergie entre l'état fondamental et l'état excité, qui, dans le cas d'un spin, est contrôlé par le champ magnétique qui est appliqué localement au spin. Le deuxième type d'action est le couplage à deux types de bains thermiques : un bain froid à température $T_c$ et un bain chaud à température $T_h$. Le spin (ou qubit) est schématisé (voir figure 1 ci-dessous) par deux barres horizontales représentant elles-mêmes le niveau d'énergie fondamental et le niveau d'énergie excité. Les sphères représentent ce qu'on appelle les populations, c'est-à-dire la probabilité que le spin soit dans le niveau d'énergie associé. Plus la sphère est importante, plus la probabilité est grande, et inversement.

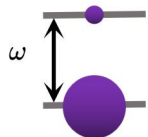

Figure 1 — **Représentation schématique d'un qubit.** Les barres horizontales représentent les niveaux d'énergie, séparées ici par un gap d'énergie ω, et les sphères représentent la probabilité que le qubit soit dans le niveau énergie correspondant.

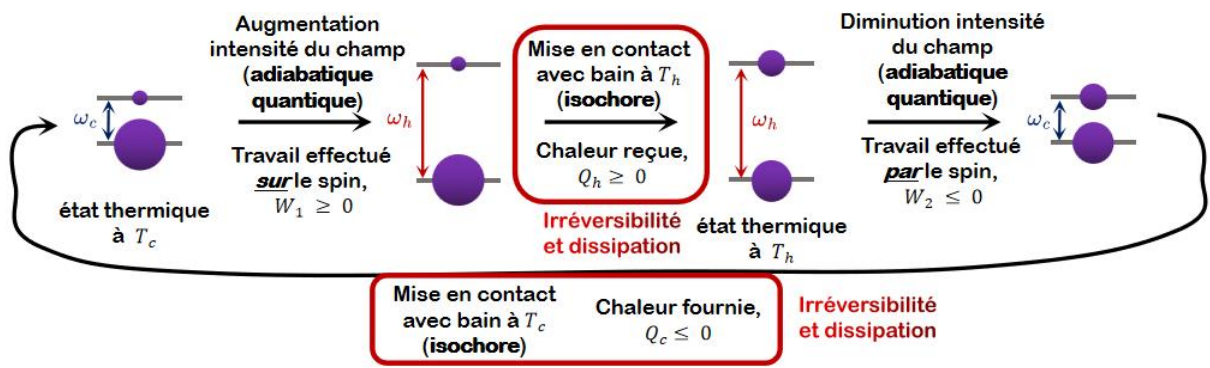

Figure 2 — **Schéma du cycle d'Otto quantique.** Voir explications dans le texte.

Le cycle, représenté dans la figure 2, commence avec le spin (ou qubit) dans un état thermique à température $T_C$ avec un gap d'énergie $\hbar\omega_c$ (où $\hbar$ est la fameuse constante de Planck, considérée égale à 1 dans les figures par soucis de simplicité). La première étape consiste à augmenter le gap d'énergie jusqu'à $\hbar\omega_h$ grâce à une augmentation de l'intensité du champ magnétique, ce qui correspond à injecter une quantité de travail $W_1$ déterminée par la différence d'énergie entre l'état à la fin de l'étape et l'état initial. La deuxième étape consiste à coupler le spin au bain chaud à température $T_h$ et le laisser thermaliser, tout en maintenant fixe le champ magnétique (et par conséquent le gap d'énergie). Cette étape génère un échange de chaleur $Q_2$ égale à la variation d'énergie durant l'étape, qui est déterminée par le changement de populations générées par l'action du bain (et représenté dans la figure 2). La troisième étape consiste à diminuer l'intensité du champ magnétique appliqué pour retourner à la valeur initiale du gap d'énergie, $\hbar\omega_c$. De manière similaire à la première étape, le travail échangé $W_2$ est égal à la variation d'énergie du spin durant l'étape. Contrairement à la première étape, le travail échangé est négatif, ce qui veut dire que le spin fourni du travail au champ magnétique. Finalement, la quatrième étape est la mise en contact avec le bain froid, suivit de la thermalisation du spin à la température $T_c$. Cette étape est associée à un échange de chaleur $Q_4$ qui est aussi négative, c'est-à-dire que le spin fourni de l'énergie au bain froid, donnée encore par la variation d'énergie du qubit durant l'interaction. L'état du spin à la fin de cette quatrième étape est un état thermique à la température $T_c$, qui est précisément égal à l'état initial : le cycle est bouclé et on peut recommencer le cycle d'opérations pour extraire plus de travail. À noter que le cycle conduit effectivement à une extraction de travail seulement lorsque le travail $W_2$ fourni par le spin et plus grand en valeur absolue que le travail $W_1$ injecté au cours de la première étape. Ceci implique la condition d'extraction de travail : $-W_1 - W_2 \geq 0$.

Quand toutes les contributions énergétiques sont sommées, on obtient zéro, $W_1 + Q_h + W_2 + Q_c = 0$, ce qui exprime le fait que le cycle est fermé. Quant à l'efficacité, elle est définie de manière traditionnelle, c'est-à-dire le travail extrait, $|W_{ext}| = -W_1 - W_2$ divisé par le coût énergétique, qui ici est la quantité d'énergie fournie par le bain chaud $Q_h$, $\eta_{Otto} = \frac{-W_1 - W_2}{Q_h}$. Après calculs explicites des $W_i$ et $Q_i$, on obtient $\eta_{Otto} = 1 - \frac{\omega_c}{\omega_h}$, qui correspond à l'efficacité du cycle d'Otto classique.

On peut observer que l'efficacité d'Otto est en fait toujours plus petite que la fameuse efficacité de Carnot, $\eta_{Carnot} = 1 - \frac{T_c}{T_h}$. Cela découle de la condition d'extraction de travail, $-W_1 - W_2 \geq 0$, qui implique $\frac{\omega_c}{\omega_h} \geq \frac{T_c}{T_h}$ et par conséquent l'inégalité $\eta_{Otto} \leq \eta_{Carnot}$. La raison physique derrière cette perte d'efficacité vient de la présence de dissipation durant les deux étapes en interaction avec les bains chaud et froid. En effet, juste avant la première étape du cycle, le qubit n'est pas dans un état d'équilibre par rapport à la température $T_h$ du bain chaud. Par conséquent, lorsque le qubit est mis en contact avec le bain chaud, la situation de non-équilibre génère de la production d'entropie et se traduit par de la dissipation et perte d'efficacité. La production d'entropie générée s'exprime comme [11, 12] $\Sigma_2 = \Delta S_2 - Q_h / T_h$, où $\Delta S_2$ désigne la variation d'entropie de von Neumann du qubit durant l'étape 2. Rappelons que pour un état quantique ρ, l'entropie de von Neumann est définie par $S(\rho) = -\text{Tr}[\rho \ln(\rho)]$. Il est aussi important de noter que l'expression [11, 12] $\Sigma_2 = \Delta S_2 - Q_h / T_h$ montre que l'origine quantique du caractère irréversible et dissipatif de la dynamique provient de la génération de corrélations entre le système quantique et son environnement.

Le même phénomène de dissipation a lieu pour l'étape en contact avec le bain froid, générant une production d'entropie égale à $\Sigma_4 = \Delta S_4 - Q_c / T_c$. À noter qu'on peut montrer que le travail extrait durant un cycle d'Otto correspond précisément au travail extrait durant un cycle de Carnot moins la quantité $(T_h \Sigma_2 + T_c \Sigma_4)$ : la production d'entropie correspond bel et bien à une perte de travail et d'efficacité.

*2.4 Le cycle de Carnot quantique*

Le cycle de Carnot quantique, représenté figure 3, peut être vu comme un raffinement du cycle d'Otto, dans lequel les sources d'irréversibilité et de perte d'efficacité identifiées dans le cycle d'Otto sont évitées. Ces sources d'irréversibilité sont évitées en faisant en sorte que le spin (qubit) soit dans un état d'équilibre thermique avec le bain chaud avant leur mise en contact. Il faut pour cela augmenter l'intensité du champ magnétique jusqu'à ce que le gap d'énergie soit égal à $\hbar\omega'_h = \frac{T_h}{T_c}\hbar\omega_c \geq \hbar\omega_h$. De manière similaire, pour éviter les sources d'irréversibilité avec l'étape en contact avec le bain froid, il faut, durant la troisième étape, diminuer l'intensité du champ magnétique jusqu'à obtenir un gap d'énergie égal à $\hbar\omega'_c = \frac{T_c}{T_h}\hbar\omega_h \leq \hbar\omega_c$. La deuxième différence avec le cycle d'Otto est durant les étapes en contact avec les bains thermiques : le gap d'énergie du spin n'est plus maintenu constant mais réduit (augmenté) durant l'étape en contact avec le bain chaud (froid), de manière à obtenir à la fin de l'étape une augmentation (diminution) significative de population excitée. Attention, pour maintenir le caractère réversible du cycle, il faut impérativement que la diminution (l'augmentation) du gap d'énergie durant l'interaction avec le bain chaud (froid) se fasse de manière quasi-statique, de sorte que le qubit soit à tout instant en équilibre thermique avec le bain chaud (froid). De cette manière, ces étapes constituent des *isothermes* (comme pour le cycle de Carnot classique).

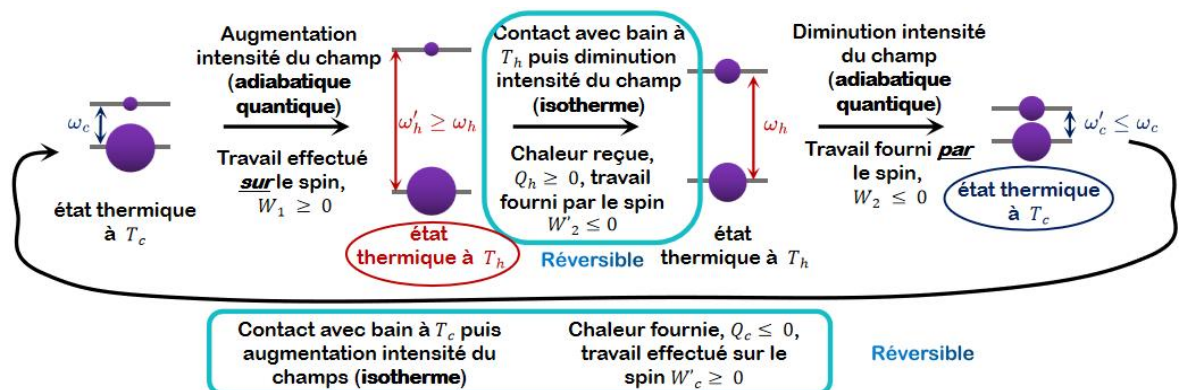

Figure 3 — **Schéma du cycle de Carnot quantique.**

En prenant en compte toutes ces caractéristiques, on peut montrer que l'efficacité du cycle de Carnot quantique est précisément égale au cycle de Carnot classique, $\eta_{\text{Carnot}} = 1 - \frac{T_c}{T_h}$. On peut d'ailleurs montrer aussi que c'est l'efficacité maximale accessible par un moteur quantique si aucune ressource quantique n'est utilisée en plus des bains thermiques chaud et froid. C'est un résultat assez fascinant : l'efficacité de Carnot, initialement établie pour des machines thermiques des siècles passés, est aussi l'efficacité maximale atteignable par des systèmes quantiques ! C'est un élément supplémentaire qui renforce l'idée d'universalité de la thermodynamique.

*2.5 Réalisations expérimentales*

Plusieurs réalisations expérimentales de machines thermiques quantiques ont déjà vu le jour. Nous en mentionnons ici seulement quelques-unes. Pour commencer, les modèles de moteurs thermiques décrits dans les deux sections précédentes ont déjà été réalisés expérimentalement en utilisant des spins nucléaires (résonance magnétique nucléaire [13]) ou des spins électroniques [14]. Mentionnons aussi que la première réalisation expérimentale utilisa un ion piégé dans un piège harmonique magnétique [15]. Bien d'autres implémentations expérimentales ont été réalisées depuis, utilisant aussi d'autres systèmes quantiques comme des atomes à deux niveaux [16] ou des centres NV (centre azote-lacune) [17].

*2.6 Effets quantiques dans les moteurs thermiques quantiques*

Abordons maintenant une question centrale : en fin de compte, les machines thermiques quantiques sont-elles les copies conformes de leurs sœurs classiques ? N'y aurait-il pas des effets quantiques à exploiter ? De nombreuses études se sont penchées sur la question et leur conclusion est sans appel : oui, il y a bien des effets quantiques. Citons pour commencer l'exemple du bain quantique contenant des cohérences quantiques [18]. Plus précisément, il s'agit d'un bain contenant des atomes à trois niveaux d'énergie. Avant d'interagir avec le moteur thermique, ces atomes sont préparés de manière identique dans un état de superposition cohérente entre deux niveaux d'énergie. Cette cohérence quantique a pour effet d'augmenter la température apparente du bain [19], ce qui conduit à une augmentation de l'efficacité du moteur [18]. À noter qu'une preuve de principe de ce moteur a récemment été réalisée expérimentalement avec une cavité optique [20]. Mentionnons aussi un mécanisme similaire dans lequel le bain chaud contient des paires d'atomes intriqués [21].

Il existe aussi un phénomène similaire pour des bains dits « bosoniques » (constitués d'ensemble d'oscillateurs harmoniques) représentant l'environnement vibrationnel ou électromagnétique d'une particule ou d'un système quantique. Ce type de bains peut aussi contenir des cohérences quantiques (états cohérents ou états « comprimés ») qui se traduisent eux aussi par une augmentation de l'efficacité du moteur [22, 23].

Une autre classe importante de machines thermiques est celle reposant sur des effets collectifs. Dans ce type de machines, au lieu d'avoir un seul qubit comme dans le cycle d'Otto et de Carnot décrit plus haut, il y a un ensemble de qubits (ou d'autres types de systèmes quantiques), et sous l'effet d'interférences constructives reliés au phénomène de superradiance, la puissance de la machine se trouve largement augmentée [24, 25].

Nous attirons maintenant l'attention du lecteur sur le point suivant : la présence d'effet quantique ne signifie pas nécessairement l'existence d'avantages quantiques. Il y a d'ailleurs un vif débat autour de la définition même d'avantages quantiques. En particulier, dans les exemples mentionnés plus haut, les performances des moteurs quantiques sont accrues grâce à l'utilisation de ressources quantiques. D'une certaine manière, il est assez naturel d'obtenir de meilleures performances si on utilise plus de ressources. Un véritable avantage quantique serait donc quand la machine quantique atteint des performances supérieures à sa sœur classique tout en utilisant les mêmes ressources. L'exemple des effets collectifs semble satisfaire une telle définition. Mentionnons aussi une étude récente [26] qui pointe à un clair avantage quantique en termes de fluctuations : un moteur quantique (continue) utilisant les mêmes ressources qu'un moteur classique peut présenter une réduction significative des **fluctuations** des courants de sortie et donc du travail extrait. C'est un avantage important puisque toute utilisation pratique requiert une grande stabilité (peu ou pas de fluctuations, voir aussi section 3). À noter que cette propriété de réduction des fluctuations est reliée à la génération de cohérences quantiques durant l'utilisation du moteur.

*2.7 Réalisations expérimentales prometteuses pour les technologies quantiques*

Indépendamment de possibles avantages quantiques, plusieurs réalisations expérimentales à fort potentiel technologique commencent à voir le jour. C'est le cas par exemple de la preuve de principe expérimentale d'un réfrigérateur quantique autonome capable de réinitialiser (c'est-à-dire refroidir) un qubit de manière compétitive avec les techniques derniers cris actuelles. Cette expérience [27] a été réalisée avec des circuits quantiques (circuits électriques contenant des composants supraconducteurs) et parvient à refroidir de manière autonome, c'est-à-dire sans contrôle extérieur actif, un qubit à des niveaux beaucoup plus faibles et de manière beaucoup plus rapide que la thermalisation naturelle du système quantique (population excitée 40 fois plus petite en un temps 67 fois plus court que la thermalisation naturelle). Le caractère autonome de cette opération représente une perspective très intéressante puisque se libérer, même partiellement, de contrôles externes actifs sur les systèmes quantiques signifie une réduction du bruit et des perturbations extérieures.

*3. Thermodynamique Quantique et Information Quantique*

Dans cette section, nous mentionnons très brièvement la question des fluctuations quantiques (quelquefois appelées « bruit » quantique). Les fluctuations ont pour origine la nature intrinsèquement probabiliste de la mesure quantique (opération quantique décrivant l'observation des systèmes quantiques) ainsi que l'interaction, inévitable, avec des systèmes externes (autres particules ou molécules, et environnement électromagnétique). Ces fluctuations quantiques peuvent changer drastiquement la dynamique d'un système quantique par rapport à sa dynamique moyenne. De cette manière, il n'est souvent pas satisfaisant de connaître seulement la dynamique moyenne, et l'étude des fluctuations devient primordiale.

De nombreuses études se penchent sur ces questions, autour par exemple des « relations d'incertitudes quantiques » (de l'anglais Thermodynamic Uncertainty Relation, TUR) [28]. Terminons en citant un des résultats les plus fascinants de ces dernières années, le théorème des fluctuations, d'abord démontré pour des systèmes classiques [29], puis étendu aux systèmes quantiques [30]. Ce théorème impose que les fluctuations d'un système classique ou quantique arbitrairement loin de l'état d'équilibre satisfont une relation d'égalité liée à la variation d'énergie libre d'équilibre du système. Autrement dit, les fluctuations ne peuvent pas être totalement arbitraires. Ce surprenant théorème a été vérifié expérimentalement pour des systèmes classiques [31, 32] ainsi que pour des systèmes quantiques [33].

### 4. La thermodynamique quantique, guide théorique pour les Technologies Quantiques

Dans cette dernière section, nous abordons les liens, de plus en plus nombreux et de plus en plus fort entre la Thermodynamique Quantique et le développement des technologies quantiques.

Plusieurs résultats théoriques récents de la Thermodynamique Quantique visent à comprendre les limites de performances de diverses opérations quantiques fondamentales pour les technologies quantiques. Ces opérations consistent par exemple à préparer le système quantique considéré dans un état particulier, ou bien induire une dynamique bien précise, ou finalement mesurer (ou observer) ce système quantique. Ces résultats récents incluent les théorèmes de vitesse limite quantique [34], qui établissent la vitesse maximale autorisée par les lois de la Mécanique Quantique à laquelle une opération donnée peut être réalisée. De plus, cette vitesse maximale est reliée aux ressources énergétiques utilisées : plus ces ressources énergétiques sont importantes, plus la vitesse maximale est élevée. À noter que la question de la vitesse des opérations est centrale dans les applications quantiques puisqu'il faut aller plus vite que le temps dit de décohérence du système au bout duquel les propriétés quantiques et l'information quantique sont perdues.

D'autres résultats étudient les coûts énergétiques de la mesure quantique, opération quantique qui décrit l'observation (indirecte) des systèmes quantiques [35,36]. Par rapport à la question énergétique, centrale dans la situation mondiale actuelle, mais aussi pour la viabilité de certaines technologies quantiques, il est important de souligner une initiative française, la « Quantum Energy Initiative » [37, 38], qui rassemble plusieurs spécialistes du domaine quantique et technologique pour étudier de manière approfondie le coût énergétique des futures technologies quantiques [38], dans le but de contribuer à les optimiser du point de vue énergétique.

Cet ensemble de résultats et d'études, appuyé par des techniques de contrôle quantique, notamment les « raccourcis à l'adiabaticité » (de l'anglais Shortcut-to-adiabaticity) [39] et le contrôle quantique optimal [40], constitue une large boîte à outils théorique et pratique pour guider le développement des futures applications et technologies quantiques. Cette boîte à outils est en pleine expansion et est amenée à être complétée par de nombreux résultats sur l'optimisation systématique des performances et du coût énergétique du contrôle quantique, ainsi que le développement de techniques adaptées aux systèmes ouverts.

### 5. Conclusions

Il existe une recherche très active autour des machines thermiques quantiques pour des applications dans des dispositifs quantiques (par exemple réinitialisation de qubits). À un niveau plus fondamental, la compréhension et le contrôle des fluctuations quantiques sont une étape nécessaire pour la réalisation de nombreuses applications et technologies quantiques. Nous avons vu aussi l'importance de comprendre et optimiser le compromis entre performances des opérations quantiques et leurs coûts énergétiques, qui est l'objet actuellement d'une intense recherche, renforcée aussi par la « Quantum Energy Initiative » [37].

En résumé, nous pouvons conclure que la Thermodynamique Quantique se veut de jouer pour les technologies quantiques le rôle qu'à jouer la thermodynamique dans la révolution industrielle : optimiser les ressources disponibles pour atteindre les meilleures performances possibles, dans la lignée des travaux de Sadi Carnot [3].